\documentclass[amsmath,preprint,amssymb]{revtex4-1}

\usepackage{graphicx}

\usepackage{dcolumn}

\usepackage{bm}

\begin{document}

\preprint{AIP/123-QED}

\title{Pre-determining the location of electromigrated gaps by nonlinear optical imaging}

\author{M.-M. Mennemanteuil}

\author{J. Dellinger}

\author{M. Buret}

\author{G. Colas des Francs}

\author{A. Bouhelier}

 \email{alexandre.bouhelier@u-bourgogne.fr}

\affiliation{Laboratoire Interdisciplinaire Carnot de bourgogne CNRS-UMR 6303, Universit\'e de Bourgogne, 21078 Dijon, France}

\date{\today}

\begin{abstract}

In this paper we describe a nonlinear imaging method employed to spatially map the occurrence of constrictions occurring on an electrically-stressed gold nanowire. The approach consists at measuring the influence of a tightly focused ultrafast pulsed laser on the electronic transport in the nanowire. We found that structural defects distributed along the nanowire are efficient nonlinear optical sources of radiation and that the differential conductance is significantly decreased when the laser is incident on such electrically-induced morphological changes.  This imaging technique is applied to pre-determined the location of the electrical failure before it occurs. 

\end{abstract}

\pacs{Valid PACS appear here}




\maketitle

Electromigration (EM) is a phenomenon involving the migration of atoms in a conductive material through the passage of a large current density \cite{Durkan1999,Lambert2003,Trouwborst2006}. This electric stress causes a momentum transfer from electrons to atoms and induces a mass transport eventually leading to the failure of electrical circuit. Widely used in molecular electronics, a controlled electromigration process provides sub-nanometer gaps where single molecule~\cite{Park1999,Mahapatro2006,Molen2003,Strachan2005} or metal particle~\cite{Lambert2003,Bolotin2004} can be inserted for single-electron tunneling experiments. These tiny gaps are also raising significant interest in the context of strongly-coupled wire optical antennas where quantum effects dominate the electromagnetic responses~\cite{aizpuruaNC12,baumberg12,dionneNL13,LiACS13} and photons can be rectified~\cite{ward10,Stolz2014}.

\begin{figure}[t]

\includegraphics[width=12cm]{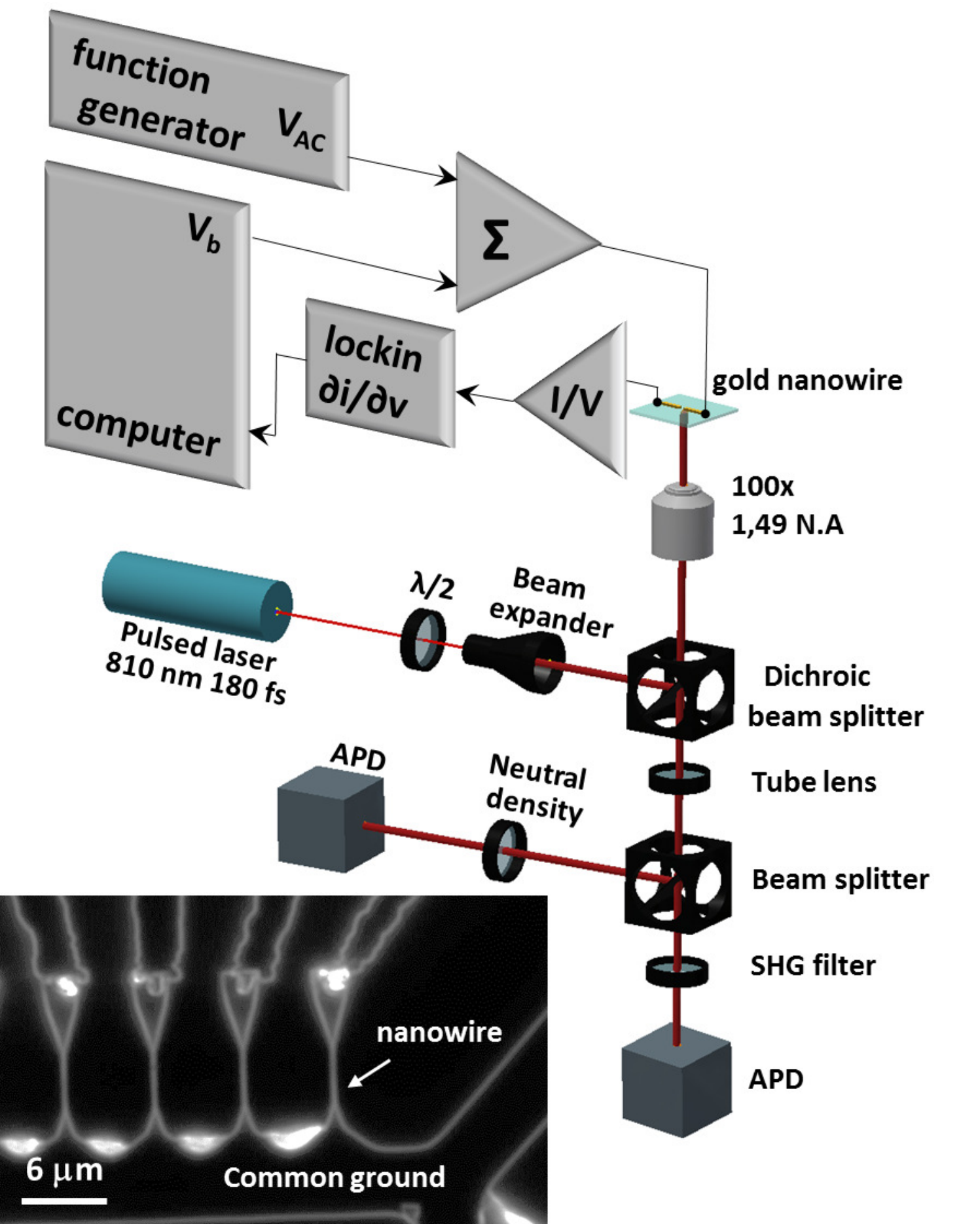} 

\caption{Schematic representation of the setup used to measure the responses of a nanowire electrically and optically stressed. A computer-controlled source is summed to an alternating voltage to initiate the electromigration of the nanowire while measuring its differential conductance. The nanowire is simultaneously raster scanned through a tightly focused ultrafast laser beam. Two avalanche photodetectors are recording the nanowire optical nonlinear responses (second-harmonic generation and two-photon luminescence). Inset: dark-field image of the Au nanowires and their electrical connections. }

\label{setup}

\end{figure}

Restructuring of an electrically-biased nanowire is a heat-assisted process where the thermal runaway typically forms large gaps~\cite{Lambert2003,Strachan2005,Trouwborst2006}. Many groups use a computer controlled feedback to prevent this catastrophic Joule heating~\cite{Strachan2005, Wu2007,Hoffmann2008,Umeno2009,Esen2009}. While the size of gap can be evaluated by monitoring the constriction resistance~\cite{Molen2003}, its location along poly-crystalline nanowire is not precisely known~\cite{Durkan1999}. In-situ  atomic force microscopy~\cite{Girod2012} and electron microscopies~ \cite{Heersche2007,Taychatanapat2007} revealed that electromigration typically takes place at grain boundaries, physical defects or impurities inherent to poly-crystalline nanowires or inevitably introduced during the  fabrication process. These pre-existing current-crowding points lead to a local rise of the temperature and are the starting point of the rupture~\cite{Ralls1989,Xuan2006}.  For chemically synthesized metal nanowires, the position of the failure is often at the electrical contact with the electrodes~\cite{Song13nanotech}.

In this work, we show that the location of the electromigrated gap can be reliably determined before the rupture of the nanowire. We developed a procedure to map the appearance of defects during the electromigration eventually leading to the formation of a gap. This technique consists at measuring the differential conductance of a biased nanowire under the influence of a pulsed laser beam. We found that the differential conductance is locally reduced when the laser irradiates  electrically-induced constrictions created along the nanowire. These specific points are spatially correlated with an enhanced nonlinear optical responses taking the form of second harmonic generation (SHG) and two-photon luminescence (TPL). We found that the defect providing the largest contrasts is systematically anticipating the location of the failing point. 

Nanowires and their electrical connections are realized by a double-step lithography followed by a conventional lift-off procedure. Macroscopic Au electrodes and a series of alignment marks are first produced by standard optical lithography. The Au nanowires are subsequently fabricated by electron beam lithography. The entire structure consist of a 3 nm thick Ti adhesion layer followed by the thermal evaporation of a 50~nm thick Au layer. The nanowires have a length of 4~$\mu$m and a width of 200~nm.  An optical dark-field image of a typical sample consisting of 24 nanowires connected to a common ground is illustrated in the inset of Fig.~\ref{setup}.

\begin{figure*}

\includegraphics[width=18cm]{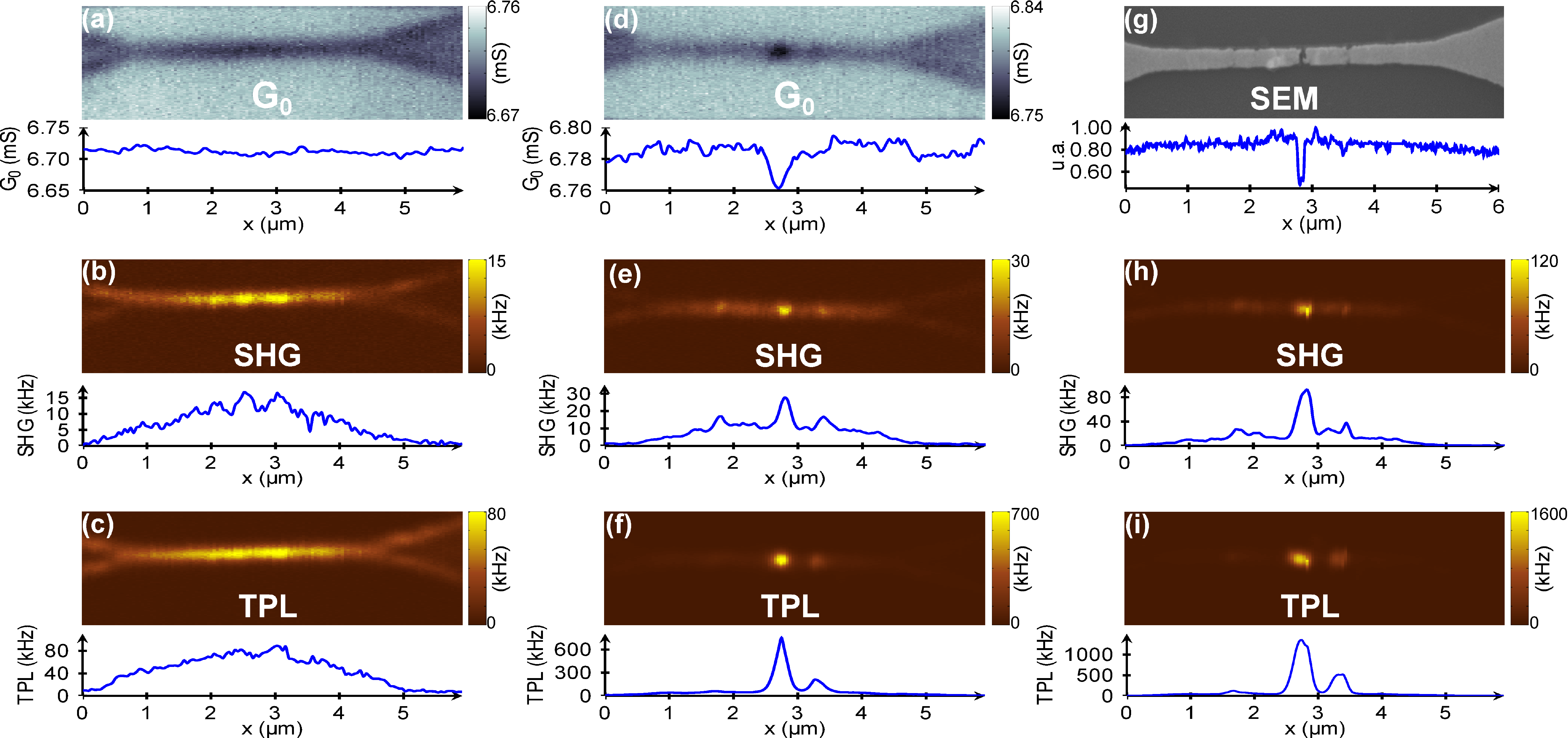} 

\caption{Characterization of electron transport and nonlinear optical responses of the nanowire after several electromigration steps. (a) Map of differential conductance  $G_{\rm o}$  before stressing the nanowire electrically. The conductance is minimal when the laser is focused on the nanowire. (b) and (c) are the simultaneously acquired SHG and TPL signals generated by the gold nanowire, respectively. (d) Map of differential conductance taken after ramping the bias several times. The darker spot is associated with the occurrence of a constriction along the nanowire. (e) and (f) are the corresponding SHG and TPL images. The defect is producing enhanced nonlinear responses. (g) SEM image of the nanowire after its electromigration. (h) and (i) are SHG and TPL images taken after the electrical rupture. An enhancement of SHG and TPL signals is generated in the gap. }

\label{confocal}

\end{figure*}

A schematic of the experimental setup is depicted in the main frame of Fig.~\ref{setup}. The nanowires are individually connected to a computer-controlled voltage source. A small modulated bias $V_{\rm mod} = V_{\rm AC} \cos(2\pi F_{\rm mod}t)$  is added to a static voltage $V_{\rm b}$ with  $V_{\rm AC}$=20~mV and $F_{\rm mod}$=1~kHz.  The alternating voltage is used to measure the total differential conductance (differential conductance of the nanowire plus that of the connexions). $V_{\rm b}$ is applied to electrically stress the nanowire. The current flowing through the nanowire is measured by a home-made current-to-voltage converter with a gain of 10$^3$~V/A that feeds a lockin amplifier referenced at $F_{\rm mod}$. The output of the locking provides a voltage signal proportional to the amplitude of the modulated current $I(F_{\rm mod})=V_{\rm AC}(\partial I_{\rm b}/\partial V_{\rm b})$. The differential conductance $G$ is then estimated by normalizing the signal by $V_{\rm AC}$.

Electromigration of the nanowire is carried out by applying a succession of voltage ramps at a rate of 50~mV/s repeated 3 times. The maximum voltage reached by the ramp, $V_{\rm F}$,  is incremented by 50~mV after each ramping sets. A return to zero between each ramps slows down the electromigration process.

After each increment, the nanowire is raster scanned through the diffraction-limited focus of an ultrafast laser emitting 180~fs pulses at a repetition rate of 80~MHz centered at a wavelength of 810 nm. Au nanostructures are known to generate various nonlinear effects when irradiated by femtosecond pulses~\cite{palomba09b,NovotnyNLbook}. Among them, second-harmonic generation (SHG) and interband-mediated two-photon luminescence (TPL) are especially strong at defect sites~\cite{Chen81,boyd86,bouhelier03,Bozhevolnyi05}. Major electrically-induced morphological modifications of the nanowire are thus expected to provide large local nonlinear responses. The position of optically active structural defaults are determined by simultaneously recording the SHG and TPL responses by two avalanche photodiodes (APD) as shown in Fig.~\ref{setup}.

The influence of the laser excitation on the differential conductance of an as-fabricated pristine nanowire is illustrated in the confocal image of Fig.~\ref{confocal} (a).  The map is constructed by  recording  the zero-bias differential conductance $G_{\rm o} (V_{\rm b}=0)$  for each position of the nanowire with respect to the focused excitation. The average laser intensity at the focus is estimated at 1.25~MW/cm$^2$.   When the nanowire or the contacting leads are situated away from the laser focus (brighter areas of Fig.~\ref{confocal} (a)) the total differential conductance is constant at  $G_{\rm o}=6.747 \pm 0.007$~mS.  When the ultrafast laser is incident on the metal structure, we observe a reduced differential conductance giving rise to the contrast of the image. The nanowire and its connecting tapers are readily recognized. The decrease of $G_{\rm o}$ is understood as a reduction of the electron mobility subsequent to the absorption of photon energy and heating of the metal lattice by electron-phonon collisions. The steady-state temperature rise is estimated at 3.9~K by using the calibration procedure described by Stolz \textit{et al.}~\cite{Stolz2014}. Figure~\ref{confocal}(b) and (c) show confocal maps of the simultaneously recorded SHG and TPL activities, respectively. The polarization of the laser is perpendicular to the axis of the nanowire. As expected, the nonlinear responses are emitted at the edges of the tapered electrical connections and along the nanowire~\cite{Berthelot:12OPEX} where the symmetry is broken~\cite{Canfield2006} and the electromagnetic field is enhanced. For the laser intensity used here, the efficiency of the TPL signal exceeds 6 orders of magnitude that of the SHG. The SHG and TPL images indicate that this pristine nanowire does not carry any strong optically-active sites. Minor variations of the nonlinear signals observed on the signal profiles originate from irregularities inherent to the nanofabrication procedure.

\begin{figure}[h]

\includegraphics[width=14cm]{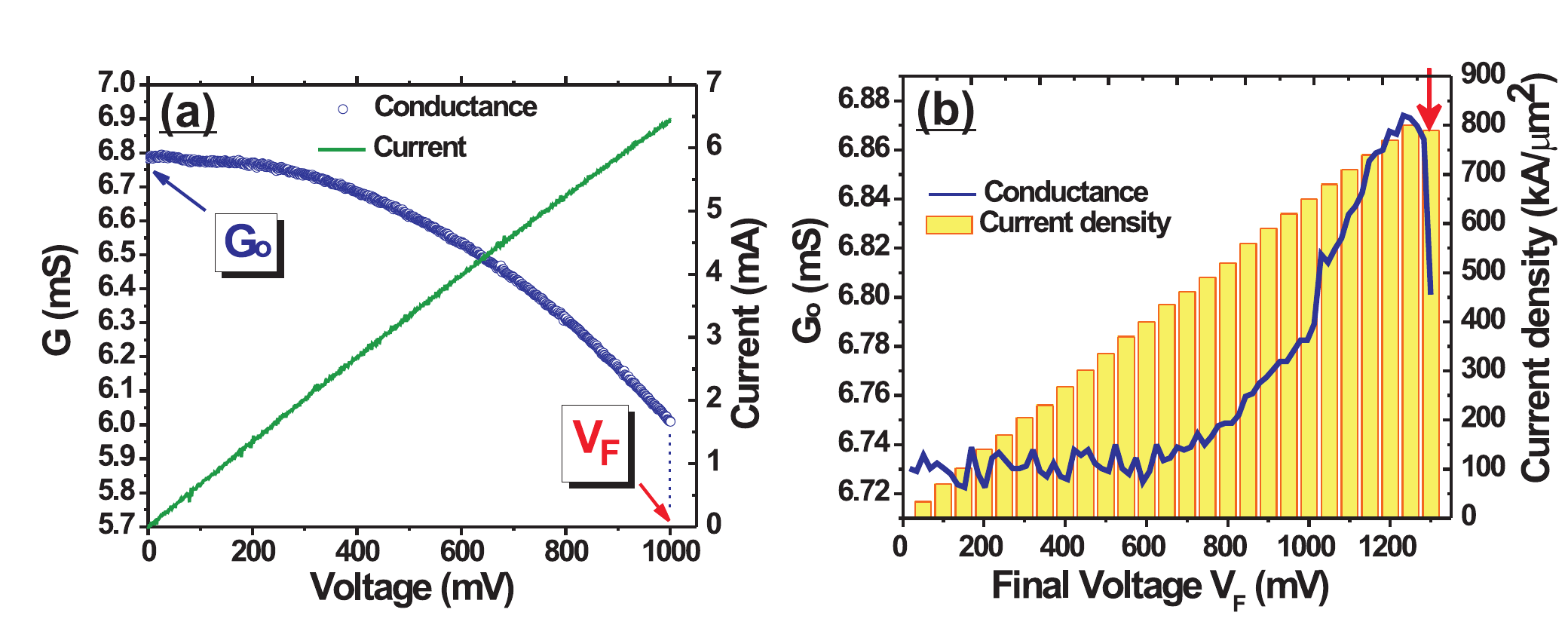}

\caption{(a) Evolution of the nanowire differential conductance $G$ and of the flowing current during a voltage ramp reaching $V_{\rm F}$=1000~mV. Joules heating is responsible for the continuous drop of $G$.  $G_{\rm o}$ is the differential conductance measured at $V_{\rm b}=0$~V. (b) Evolution of $ G_{\rm o}$ and the current density with the highest voltage reached during ramping. The arrow indicates the transition voltage where $G_{\rm o}$ inflects and the electrical conditions corresponding to Fig.~\ref{confocal}(d) to (f).}

\label{conductance}

\end{figure}

Electromigration of the nanowire is then initiated by successively ramping the static voltage $V_{\rm b}$.  Figure~\ref{conductance}(a) shows an example of the evolution of the conductance $G$ and the current flowing in the nanowire during a voltage ramp. The final ramping value is $V_{\rm F}$=1000~mV. The current-to-voltage characteristic starts to deviate from a linear relationship at very low bias as demonstrated by the monotonous decrease of $G$ after $\sim$ 70~mV. This conductance drop results from the rise of the temperature when the current is flowing and dissipates energy in the nanowire~\cite{dumpich07}. We previously determined the rate of variation at -0.0147~mS/K relatively to the ambient temperature~\cite{Stolz2014}. The temperature rise of the nanowire at $V_{\rm b}$=$V_{\rm F}$ is thus estimated at $\Delta T$=52~K. For consistency check, we also determined the linear temperature coefficient $\alpha=(G_{\rm o}/G_{\rm F}-1)\Delta T^{-1}$ where $G_{\rm F}$ is the differential conductance at $V_{\rm F}$. We find $\alpha$=0.0024 K$^{-1}$ in agreement with the value measured by Stahlmecke and Dumpich for similar nanowires~\cite{dumpich07}.

The evolution of zero-bias conductance $G_{\rm o}$ with the final ramping bias is a good indicator to monitor the onset of  electromigration. Figure~\ref{conductance}(b) illustrates how $G_{\rm o}$ changes when $V_{\rm b}$ is successively ramped to higher voltage $V_{\rm F}$.  $G_{\rm o}$, measured at the beginning of each ramp, stays constant until $V_{\rm F}\sim 700$~mV corresponding to a current density of 430~kA/$\mu$m$^2$. After this threshold, the electrical performance of the structure is  improved as shown by the steady increase of $G_{\rm o}$ between 700~mV$<V_{\rm F}<$1250~mV.  At this voltage, the temperature of the nanowire is estimated at 318~K. This modest temperature is not sufficient to provoke an annealing of the poly-crystalline Au nanowire~\cite{Potma13}.  The increase of  $G_{\rm o}$ probably results from the thermal desorption of nanowire adsorbates (\textit{e.g.} water layer). We confirmed this hypothesis by measuring $G_{\rm o}$ again after 96 hours and found the same characteristic rise of the zero-bias conductance indicating that the process is reversible. 
 
 When the current density reaches 800kA/$\mu$m$^2$, the zero-bias differential conductance suddenly drops. This inversion point (arrow) is a clear signature of the formation of irreversible electromigration-induced defects along the nanowire. Under this electrical regime, the variation of $G$ with voltage indicates a nanowire temperature of 418~K, a temperature typical for triggering the electromigration process~\cite{Wu2007,Hoffmann2008}.

We confirmed the presence of voids after the inversion point of $G_{\rm o}$ (arrow in Fig.~\ref{conductance}) by subsequently imaging the nanowire zero-bias differential conductance, the SHG and TPL signals as a function of nanowire position with respect to the focus of the ultrafast laser. The image of Fig.~\ref{confocal}(d) shows the laser influence on conductance map. Compared to the pristine nanowire [Fig.~\ref{confocal}(a)], we observe a localized region at the center of the nanowire where the laser causes a larger decrease of  $G_{\rm o}$. This area is spatially correlated to a strong response on the SHG and TPL images as observed in  Fig.~\ref{confocal}(e) and (f). The images also reveal satellite active sites with weaker nonlinear responses indicating that the nanowire is simultaneously thinned at competing current-crowding points. Interestingly, however, the strongest nonlinear response coincides with the largest conductance drop at the center of the nanowire. The defect acts as an entry gate for dissipating the laser energy in the nanowire. At this position the laser-induced rise of temperature remains modest at 2~K, but is sufficient to provide an imaging contrast.  After this characterization step, the electromigration of the nanowire is completed by applying a final series of ramp at  $V_{\rm F}=1300$~mV. The rupture of the nanowire is confirmed when the conductance is no longer measurable. Figure~\ref{confocal}(g) shows a scanning electron micrograph (SEM) of the nanowire after its failure. The location of the gap identified in the SEM image produces intense nonlinear activities [Fig.~\ref{confocal}(h) and (i)] at the position predicted by the largest conductance drop measured before the failure of the nanowire [Fig.~\ref{confocal}(d)]. 

We have repeated the same imaging protocol and electromigration procedure on other nanowires and systematically found that the location of the largest conductance drop observed after the onset of electromigration [arrow in Fig.~\ref{conductance}(b)] anticipates the position of the electrical failure point.

The ability to predict the spatial location of electromigrated gaps is an important step for fabricating advanced electrically-controlled optical antennas designs. Simple tunneling gaps are essential for rectifying electromagnetic radiation~\cite{ward10,Stolz2014,Elezzabi13} or producing local plasmonic sources~\cite{Bharadwaj11,Lemoal13}, but are plagued by low quantum yields. Engineering the electromagnetic landscape at the vicinity of the gap may contribute at improving the interaction cross-sections by deploying resonant feedgaps~\cite{Mayer09,HechtNL12}. However, the different nanofabrication steps that are necessary for patterning the gap with parasitic units (\textit{e.g.} Yagi-Uda elements~\cite{curto10}) may compromise the integrity of this sensitive tunneling device. Our approach provides a mean for pre-engineering the predicted location of gap before acquiring its full functionality.

The research leading to these results has received funding from the European Research Council under the European Community's Seventh Framework Program FP7/2007-2013 Grant Agreement no 306772. This project is in cooperation with the Labex ACTION program (contract ANR-11-LABX-01-01).

%

\end{document}